\documentclass[11pt,letterpaper,superscriptaddress]{JHEP3}
\usepackage{graphicx}
\usepackage{amssymb, amsmath,amsfonts,amstext,graphics}
\usepackage{epstopdf}
\usepackage{cite}
\input epsf.tex
\newcommand{\lang}{\mathcal{L}}

\newcommand{\beq}{\begin{eqnarray}}% can be used as {equation} or {align}
\newcommand{\eeq}{\end{eqnarray}}

\def\ltap{\ \raise.3ex\hbox{$<$\kern-.75em\lower1ex\hbox{$\sim$}}\ }
\def\gtap{\ \raise.3ex\hbox{$>$\kern-.75em\lower1ex\hbox{$\sim$}}\ }
\newcommand{\gsim}{\lower.7ex\hbox{$\;\stackrel{\textstyle>}{\sim}\;$}}
\newcommand{\lsim}{\lower.7ex\hbox{$\;\stackrel{\textstyle<}{\sim}\;$}}

\def\CO{{\cal O}}

\def\CO{{\cal O}}

\def\be{\begin{equation}}
\def\ee{\end{equation}}
\def\bea{\begin{align}}
\def\eea{\end{align}}

\newcommand{\MO}{\mathcal{O}}
\def\unit{\relax{\rm 1\kern-.26em I}}
\newcommand{\half}{{\frac{1}{2}}}

\newcommand{\Tr}{{\text{ Tr }}}

\preprint{SU-ITP-10-16}
\title{Simple Models of Superconformal Flavor}
\author{Nathaniel J. Craig$^a$\\ 
$^a$ Institute for Theoretical Physics, Stanford University,\\ 
Stanford, CA 94306, U.S.A. \\
E-mail: {\tt ncraig@stanford.edu}}
\abstract{ The observed hierarchy of fermion masses and mixings may be generated by renormalization group flow if the Standard Model is coupled to a near-conformal sector at high energies. If the conformal sector is supersymmetric, these effects are rendered calculable by a combination of superconformal symmetry and a-maximization. The viability of such models depends on whether they generate the observed fermion mass hierarchy before the Standard Model gauge couplings hit a Landau pole. Here we construct a variety of simple vector-like models of superconformal flavor, including both ten-centered and democratic variations. We discuss in detail the subtleties of applying the a-maximization procedure to determine anomalous dimensions of Standard Model fields. We find that a wide range of models based on SU(N) or Sp(2N) SQCD with fundamental and adjoint matter are viable theories of superconformal flavor.
}

\begin{document}
\maketitle

\section{Introduction}

	The observed hierarchy of fermion masses and mixings remains one of the most puzzling features of the Standard Model; the masses of three generations of quarks and leptons range over more than five orders of magnitude. Yet the observed spectrum is not entirely random, but rather seems to reflect an underlying structure. The masses of subsequent generations are arrayed with nearly even spacing; intergenerational mixings exhibit a nearest-neighbor hierarchy; and the similarity of down-type quark and lepton masses relative to up-type quark masses is suggestive of a grand unified theory (GUT) at high energies.
	
	 It seems likely that a comprehensive ultraviolet completion of the Standard Model may feature some explanation for this apparent flavor structure. Perhaps the most common approach to such theories of flavor involves engineering the observed Yukawa textures directly, either through approximate symmetries or radiative corrections \cite{Froggatt:1978nt, Leurer:1992wg, Leurer:1993gy, Grossman:1995hk, Dobrescu:2008sz,Graham:2009gr}. A compelling alternative approach is to treat the Yukawa matrices as entirely anarchical, consistent with effective field theory, and generate the flavor hierarchy through wavefunction renormalization. In four dimensions, this may be readily accomplished by coupling the Standard Model to a sector with strong conformal dynamics \cite{Georgi:1983mq, Nelson:2000sn, Poland:2009yb, Aharony:2010ch, Antola:2010nt} or by assembling Standard Model fermions themselves as composites of some strong dynamics \cite{ArkaniHamed:1997fq, Luty:1998vr, Franco:2009wf, Craig:2009hf}. The Yukawa matrices then acquire the desired hierarchical form in terms of the canonically-normalized low-energy degrees of freedom. In both cases, supersymmetry frequently plays a role, both in rendering calculable the strong dynamics responsible for flavor hierarchies and in explaining the scale of electroweak symmetry breaking. This raises the attractive prospect that the hierarchy problem, Standard Model flavor hierarchy, and supersymmetric flavor problem may all share a common explanation and correlated features.

	  In this paper we wish to focus on specific models of flavor produced by coupling the Standard Model to a sector with strong conformal dynamics over a range of energies in which both sectors are supersymmetric. Among other features, such models have the virtue of considerable predictivity, as the anomalous dimensions (and hence fermion masses) of Standard Model fields are determined entirely by the gauge group and matter content of the superconformal sector. The primary challenge in building such models is to explicitly determine these anomalous dimensions, which at the time of the original work in \cite{ Nelson:2000sn} was difficult to achieve for simple models without a proliferation of superpotential couplings. In \cite{Poland:2009yb}, considerable progress was made towards studying vector-like models of superconformal flavor using the $a$-maximization procedure \cite{Intriligator:2003jj} to determine the anomalous dimensions of both Standard Model and SCFT fields. Here our approach follows closely that of \cite{Poland:2009yb}, using $a$-maximization to investigate the flavor spectrum arising from simple vector-like superconformal sectors.

	In principle, this picture of flavor is related by a loose version of the AdS/CFT correspondence \cite{Maldacena:1997re,Gubser:1998bc,Witten:1998qj,ArkaniHamed:2000ds, Rattazzi:2000hs} to warped 5D models \cite{Randall:1999ee} in which each field has an exponential profile fixed by its bulk mass \cite{Gherghetta:2000qt, Huber:2000ie}. Insofar as these 5D models possess the virtues of calculability and parametric freedom, one might naturally wonder whether 4D superconformal models have any real advantages over their 5D duals. Among other things, models based on 4D CFTs have fewer free parameters, as anomalous dimensions of all fields are fixed by the superconformal algebra and marginal interactions. Although various flavor textures may be realized in 5D by adjusting bulk masses, this leads to a proliferation of parameters, and it is not entirely clear whether a given 5D theory possesses a dual 4D CFT. Moreover, as the AdS/CFT correspondence is a strictly large-$N$ duality, studying theories based on 4D CFTs at small $N$ may reveal features not readily accessible in 5D duals \cite{Roy:2007nz, Murayama:2007ge, Craig:2009rk}.

Our paper is organized as follows: In Section 2 we introduce the philosophy of superconformal flavor, relevant constraints, and previous results. In Section 3 we review the relation between $R$ charges and scaling dimensions at superconformal fixed points, as well as the $a$-maximization procedure for determining the superconformal $R$-symmetry. Some subtlety arises when gauge-invariant operators saturate the unitarity bound, which influences both the $a$-maximization procedure and the contribution of SCFT states to the Standard Model $\beta$ function.  Having established the necessary tools, we turn in Section 4 to simple models of superconformal flavor based on $SU(N)$ gauge theories with an adjoint chiral superfield. In Section 5 we briefly treat related models based on $Sp(2N)$ gauge theories, which have the virtue of significantly smaller matter content charged under $SU(5)_{SM}$. In Section 6 we discuss the fixed points of these models and issues related to decoupling. We reserve for the Appendix the detailed constraints and numerical results of the $a$-maximization procedure as applied to the models in Sections 4 and 5.

\section{Flavor hierarchy from flavor anarchy}

The essential philosophy of superconformal flavor stems from the observation that fermion mass ratios and mixings may arise in the infrared from anarchy in the ultraviolet due strictly to quantum renormalization effects \cite{Georgi:1983mq}. The size of renormalization effects required to explain the flavor hierarchy points to strongly coupled dynamics, which are in general incalculable. If, however, the dynamics are supersymmetric and approximately conformal, these strong renormalization effects may be estimated accurately \cite{Nelson:2000sn}. In this section we will first review the means of generating Standard Model flavor hierarchies through large wavefunction renormalization, before turning to the approximately superconformal sectors that may be responsible. For simplicity, and to avoid potential conflicts with experimental constraints on baryon and lepton number-violating operators, we will restrict our attention to models of superconformal flavor operating at and above the GUT scale. This simplifying assumption leads us to consider Standard Model fermions strictly as components of GUT multiplets. 

\subsection{Standard Model flavor physics}
The Standard Model Yukawa couplings are of the form (in unified notation)
\be
W_{SM} \supset y_u^{ij} T_i T_j H_u + y_d^{ij} T_i \overline{F}_j H_d 
\ee 
where the $T_i \supset Q_i, U_i, E_i$ transform as a $\mathbf{10}$ of $SU(5)_{SM}$ and the $\overline F_i \supset L_i, D_i$ transform as a $\overline{\mathbf{5}}$, with $i = 1,2,3$. (We will not consider here the source of neutrino masses, but these may be included fairly easily.) The philosophy of superconformal flavor is simply that the Yukawa matrices $\mathbf{y}_u$ and $\mathbf{y}_d$ are not intrinsically hierarchical in the far ultraviolet, but rather contain $\mathcal{O}(1)$ factors consistent with effective field theory. The observed hierarchy in the Yukawas arises because the fields $T_i$ and $\overline{F}_i$ inherit large wavefunction renormalization factors at some lower scale through coupling to an approximately conformal sector. When the infrared degrees of freedom are canonically normalized, the $\mathcal{O}(1)$ entries of the Yukawa matrices accumulate additional family-dependent suppression factors.

To see how this comes about, assume the fields $\Phi_i = T_i, \overline{F}_i$ of the Standard Model acquire large wavefunction renormalizations $Z_i(\mu)$ in the holomorphic basis at a scale $\mu \simeq M_{GUT}$:
\be
\lang = \int d^4 \theta \sum_i Z_i(\mu) \Phi_i^\dag \Phi_i
\ee
In the physical basis where fields are canonically normalized, this leads to suppression factors $\epsilon_i \equiv 1/\sqrt{Z_i}$ in the Yukawa couplings. In this notation, the Yukawa couplings are given by
\be
W_{SM} \supset \epsilon_{T_i} \epsilon_{T_j} y_{u,0}^{ij} T_i T_j H_u + \epsilon_{T_i} \epsilon_{\overline F_j} y_{d,0}^{ij} T_i \overline{F}_j H_d 
\ee 
where $y_{u,0}^{ij}, y_{d,0}^{ij} \sim \mathcal{O}(1)$ are the anarchical Yukawa coefficients from the ultraviolet theory. To within these $ \mathcal{O}(1)$ coefficients, the quark and lepton masses are therefore given by 
\begin{eqnarray}
(m_t, m_c, m_u) \approx \frac{1}{\sqrt{2}} v \sin \beta \left(\epsilon_{T_3} \epsilon_{T_3} \epsilon_{H_u} , \epsilon_{T_2} \epsilon_{T_2} \epsilon_{H_u}, \epsilon_{T_1} \epsilon_{T_1} \epsilon_{H_u} \right) \\ \nonumber
(m_b, m_s, m_d) \approx \frac{1}{\sqrt{2}} v \cos \beta  \left(\epsilon_{T_3} \epsilon_{\bar F_3} \epsilon_{H_d} , \epsilon_{T_2} \epsilon_{\bar F_2} \epsilon_{H_d} ,\epsilon_{T_1} \epsilon_{\bar F_1} \epsilon_{H_d} \right) \\ \nonumber
(m_\tau, m_\mu, m_e) \approx \frac{1}{\sqrt{2}} v \cos \beta \left(\epsilon_{T_3} \epsilon_{\bar F_3} \epsilon_{H_d} , \epsilon_{T_2} \epsilon_{\bar F_2} \epsilon_{H_d} ,\epsilon_{T_1} \epsilon_{\bar F_1} \epsilon_{H_d} \right)
\end{eqnarray}
where $v \approx 246$ GeV as usual. The resulting mixing angles in the $CKM$ matrix are
\begin{eqnarray}
|V_{CKM} | \approx \left( \begin{array}{ccc}
1 & \epsilon_{T_1}/\epsilon_{T_2} & \epsilon_{T_1}/\epsilon_{T_3} \\
\epsilon_{T_1}/\epsilon_{T_2} & 1 & \epsilon_{T_2}/\epsilon_{T_3} \\
\epsilon_{T_1}/\epsilon_{T_3} & \epsilon_{T_2}/\epsilon_{T_3} & 1 
\end{array} \right) \hspace{5mm}
\end{eqnarray}  
which offers a fairly good parameterization of the observed values. It is natural, then, to consider what values of $\epsilon_i$ are required to match the observed masses of Standard Model fermions. This is not an exact science; the $\epsilon_i$ should be chosen to produce quark and lepton masses at the GUT scale, where they are subject to potentially sizable uncertainties due to supersymmetric threshold corrections. But a reasonable estimate gives (assuming $\epsilon_{H_u} \sim \epsilon_{H_d} \sim 1$) \cite{Antusch:2008tf, Poland:2009yb}
\beq \label{flavhier}
\epsilon_{T_i} &\approx& \left( 0.001 \div 0.002, 0.03 \div 0.04, 0.7 \div 0.8 \right) \\ \nonumber
\epsilon_{\overline F_i} &\approx& \tan \beta \cdot \left(0.002 \div 0.01, 0.001 \div 0.007, 0.006 \div 0.02 \right)
\eeq
where the values of $\epsilon_{\overline F_i} $ come from considering the down-type quark masses; the result for lepton masses gives, encouragingly, similar results to within uncertainties due to threshold corrections.

There are clearly various ways to generate the hierarchy of (\ref{flavhier}). We will focus here on two types of flavor structures. The first is the so-called ``ten-centered'' structure, in which only the $T_i$ of the Standard Model obtain significant $\epsilon$ factors from the conformal sector. The utility of these models stems from the observation that Standard Model flavor looks to be driven predominantly by a hierarchy among the different generations of $T_i$. For this to work also requires a large value of $\tan \beta$, which may cause problems with proton decay. Alternatively, we will also consider more ``democratic'' models with coupling to both $T_i$ and $\overline F_i$. Here it is possible to accommodate much smaller values of $\tan \beta$, but one must be careful not to generate over-large hierarchies among the $\overline F_i$. 

\subsection{Superconformal flavor physics }

Given that the flavor hierarchy may be explained by large wavefunction renormalization of Standard Model fields, it is now a matter of determining how such large renormalization might arise. Typically, the renormalization of Yukawa couplings and other dimensionless parameters in 4D scales logarithmically with energy, which is poorly suited to generating the required large factors (up to $Z_i \sim 10^6$) required to explain the range of quark masses. The key point, however, is that such significant effects may be realized in theories with approximately scale-invariant gauge couplings. If Standard Model fields $\Phi_i$ couple over some range of energies to fields charged under an approximately conformal gauge group $G$, these couplings may generate large anomalous dimensions $\gamma_i$. In this case the wave-function renormalization of the fields $\Phi_i$ is given in terms of the anomalous dimension $\gamma_i$ by $\frac{d}{dt} \log Z_i \approx - \gamma_i$ (where $t = \log \mu$). If the group $G$ becomes approximately conformal at a scale $\Lambda$ and flows away from the conformal fixed point at a scale $\Lambda_*$, the resulting suppression factors take the form
\be
\epsilon_{\Phi_i} = \exp \left( - \half \int_{\log \Lambda_*}^{\log \Lambda} \gamma_i d t \right)
\ee
Of course, the running of Standard Model gauge and Yukawa couplings spoils conformal invariance. For an approximately superconformal fixed point, the anomalous dimensions are constant up to corrections of order $g^2_{SM} / 16 \pi^2$ and $y^2 / 16 \pi^2$. In this approximation, the $\gamma_i$ are constant, and so we find 
\be
\epsilon_{\Phi_i} = \left( \frac{\Lambda_*}{\Lambda} \right)^{\gamma_i / 2}
\ee
When the anomalous dimensions are sufficiently large -- typically $\gamma_i \sim \mathcal{O}(1)$ -- and the range of energies $\Lambda_* < \mu < \Lambda$ sufficiently long, the suppression factors required by the flavor hierarchy will be readily generated.

The large anomalous dimensions $\gamma_i$ for Standard Model fields may be generated by coupling to operators of the conformal sector via marginal interactions at the conformal fixed point. When the fixed point is superconformal, these anomalous dimensions become calculable. At such a fixed point, there is a simple relation between the scaling dimension of a chiral primary operator $\mathcal{O}$ and its superconformal $R$-charge, $\dim(\mathcal{O}) = \frac{3}{2} R_\mathcal{O}$. Correspondingly, the anomalous dimensions of operators at the superconformal fixed point are given by  $\gamma_{\mathcal{O}} = 3 R_{\mathcal{O}} - 2$. Since marginal superpotential terms at the fixed point must have $R$-charge 2, a superpotential coupling between SM and SCFT fields of the form $W = \Phi_i \MO$ implies $R_{\Phi_i} = 2 - R_\MO$ and hence $\gamma_i = 4 - 3 R_\MO$. Such couplings require $\MO$ to transform nontrivially under Standard Model gauge symmetries. We may accomplish this by weakly gauging an $SU(5)_{SM}$ subgroup of the flavor symmetries in the SCFT, which amounts to a (small) explicit breaking of the global symmetry group. One caveat of this discussion is that the correct superconformal $R$ charges may not always be readily determined, a challenge we will turn to in \S 3.

In this paper we will restrict our focus to vector-like superconformal sectors with fundamental matter and a rank-2 (adjoint or antisymmetric) tensor field. Such higher-rank fields introduce new gauge-invariant chiral operators whose canonical dimensions and $R$-charges may differ significantly from those of gauge-invariant operators formed by fundamental  fields alone. As successful models of superconformal flavor require two or more operators with the same $SU(5)_{SM}$ charges but substantially different $R$-charges, rank-2 tensor fields therefore comprise an essential ingredient of superconformal flavor engineering. This situation is highly reminiscent of 4D models of composite flavor, in which various Standard Model families arise as mesons of identical SM charges but differing canonical dimensions \cite{Luty:1998vr, Craig:2009hf}. Likewise, we focus on vector-like conformal sectors due both to their ubiquity (as such gauge sectors often arise in string compactifications) and the simplicity with which their exotic states may be decoupled.

\subsection{Constraints on superconformal sectors}

It is not quite enough to simply generate large wavefunction renormalization for Standard Model fermions. Weakly gauging a subgroup of the SCFT flavor symmetries leads to a plethora of extra states charged under both $G$ and $SU(5)_{SM}$, which must be decoupled well before the weak scale, both to avoid spoiling Standard Model gauge coupling unification and violating observational bounds on SM-charged exotics. For simplicity, we assume the decoupling occurs at a scale $\sim \Lambda_*$, due to some small relevant deformations of the SCFT. It is not necessary to assume that all SCFT states acquire masses of $\mathcal{O}(\Lambda_*)$; it may be the case that some fields survive to lower energies, and perhaps are responsible for supersymmetry breaking at a lower scale.

If Standard Model fields decouple from the superconformal sector at $\Lambda_*$, we still need to worry about irrelevant operators induced at this scale.  Foremost is the need to avoid operators violating baryon and lepton number. Although in general such operators will not be induced directly by the superpotential couplings of our theory, they are expected to appear as dimension-six operators in the K\"{a}hler potential with suppression of order $16 \pi^2 / \Lambda_*^2$. If $\Lambda_* \sim M_{GUT}$, this is fairly safe for all but the largest values of $\tan \beta$. 

\begin{figure}[t] %  figure placement: here, top, bottom, or page
   \centering
   \includegraphics[width=6in]{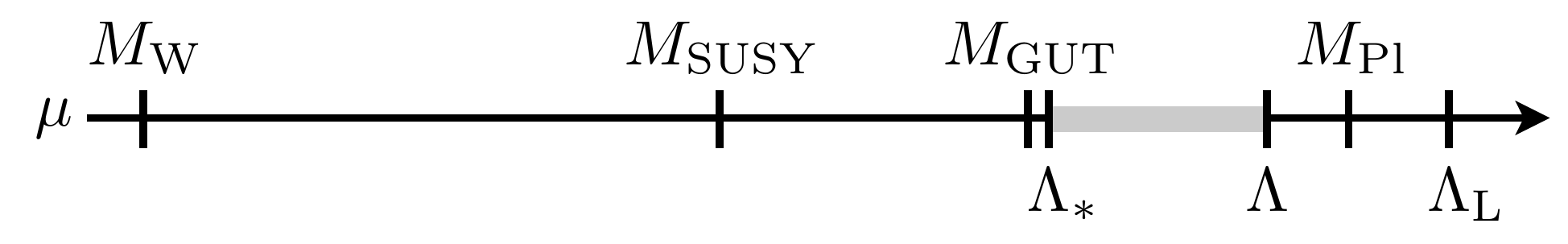} 
   \caption{Cartoon of energy scales. We assume the gauge group $G$ is approximately conformal in the energy range $ \Lambda_* < \mu < \Lambda$; that the fields charged under $G$ decouple from the Standard Model around $M_{GUT} \sim \Lambda_*$; and that any Standard Model Landau poles lie at a scale $\Lambda_L > \Lambda$ (and ideally $ > M_{Pl}$). In principle, fields charged under $G$ may be responsible for breaking supersymmetry at a scale $M_{SUSY}$.}
   \label{fig:line}
\end{figure}

Whatever the scale of superconformal flavor, a principal constraint arises from the requirement that the Standard Model gauge couplings remain perturbative long enough for the observed flavor hierarchy to be generated. Consistency of our models requires that $SU(5)_{SM}$ remain a weakly-gauged subgroup of the SCFT flavor symmetries while at the approximately conformal fixed point, but the addition of so much extra matter charged under $SU(5)_{SM}$ tends to generate a Landau pole for $g_5$. Successful model-building amounts to ensuring that the Landau pole lie at a scale $\Lambda_L$ above the window of energies in which the flavor hierarchy is produced. 

This constraint may be enforced quite easily. The NSVZ beta function  \cite{Novikov:1983uc} for the $SU(5)_{SM}$ gauge coupling $g_5$ is given by
\be
\beta_{g_5} = - \frac{g_5^3}{16 \pi^2} \frac{\left[ 15 - \sum_i T(r_i) (1-\gamma_i) \right]}{1-5 g_5^2/8 \pi^2}
= \frac{g_5^3}{16 \pi^2} \frac{ \mathfrak{b}}{1- 5 g_5^2 / 8 \pi^2} 
\ee 
where $\mathfrak{b}$ is the ``exact'' $\beta$-function coefficient
\be
\mathfrak{b} \equiv - 15 + \sum_i (1-\gamma_i) T(r_i) = - 3 \Tr \left[ U(1)_R SU(5)_{SM}^2 \right]
\ee
Given the $R$-charges of the SCFT fields charged under $SU(5)_{SM}$, we may compute $\mathfrak{b}$ for a given model and determine the scale $\Lambda_L$ at which $g_5$ hits a Landau pole (subject to some subtleties arising when operators of the SCFT sector go free, as we will discuss in the next section). It is amusing to note that this $\beta$-function coefficient is equivalent to the $ U(1)_R SU(5)_{SM}^2$ global anomaly coefficient at the conformal fixed point. This will turn out to play an important role in computing the contribution to $\beta_{g_5}$ from gauge-invariant chiral operators that saturate the unitarity bound.

Using these results, we subject the models under consideration to a fairly simple criterion: that they generate an adequate flavor hierarchy over the range $\Lambda_* < \mu < \Lambda$ smaller than the hierarchy $\Lambda_* < \mu < \Lambda_L$ between decoupling and the Landau pole for $g_5$. For all models we compute $\Lambda_L / \Lambda_*$ assuming Standard Model field content, an additional $SU(5)_{SM}$ adjoint Higgs $\Sigma$, and the field content of the superconformal sector. For ten-centered models, we compare this to the ratio $\Lambda / \Lambda_*$ required to get within a factor of $3$ of the observed hierarchy in up-type quark masses. For democratic models, we compare this to both the ratio $\Lambda_T / \Lambda_*$ required to get within a factor of 3 of the up-type quark hierarchy, and the ratio  $\Lambda_F / \Lambda_*$ required to get within a factor of 3 of the lepton mass hierarchy for $\tan \beta = 10$.

\section{R charges from $a$-maximization}

Clearly, in order for the superconformal flavor mechanism to be effective it is necessary for the conformal sector to generate sufficiently large anomalous dimensions for Standard Model fields. Ideally, these anomalous dimensions should be calculable -- often a challenging proposition for strongly coupled theories. A tremendous advantage is gained if the universe is supersymmetric over the energy range in which the flavor hierarchy is generated. In this case the superconformal algebra relates the scaling dimension of gauge-invariant chiral operators to their transformation properties under the superconformal $U(1)_R$ symmetry.

Recall that the superconformal algebra is the superalgebra $SU(2,2|1),$ the bosonic part of which consists of the familiar conformal $SO(4,2)$ and an additional non-anomalous $U(1)_R$. The charges of gauge-invariant chiral primary operators under this particular $U(1)_R$ give, in turn, their scaling dimension at the conformal fixed point. When this $U(1)_R$ may be readily identified, it provides a direct means of computing anomalous dimensions for fields coupled to the SCFT; such was the strategy employed in \cite{Nelson:2000sn}. However, the utility of this approach is limited by the ease with which the superconformal $U(1)_R$ may be identified. In general, superconformal theories possess a variety of candidate $U(1)_R$ symmetries; the principle challenge lies in determining which $U(1)_R$ dictates the scaling dimensions of chiral primary operators at the fixed point.

Once the correct $R$-charges are known, unitarity imposes a bound relating the scaling dimension of a gauge-invariant operator $\mathcal{O}$ and its $R$-charge via the inequality \cite{Minwalla:1997ka}
\be \label{eqn:dim}
\Delta (\mathcal{O}) \geq |\frac{3}{2} R(\mathcal{O}) |
\ee
The inequality is saturated for chiral and antichiral primary operators; we will henceforth be interested solely in gauge-invariant chiral primaries for which the equality in (\ref{eqn:dim}) is exact. Once we know the scaling dimension, we can express the anomalous dimension $\gamma$ of an operator in terms of its R-charges; for a chiral primary,
\be
\Delta_{\MO} \equiv 1 + \half \gamma_{\MO} = \frac{3}{2} R_{\MO} \rightarrow \gamma_{\MO} = 3 R_{\MO}- 2
\ee
Clearly, if we can compute the $R$-charges of operators under the superconformal $U(1)_R$ symmetry, we may determine (up to the usual corrections of order $g_5^2/16 \pi^2$ coming from the gauging of flavor symmetries) the anomalous dimensions of fields in a given model of flavor anarchy. But therein lies the rub; in general, a given theory will possess a variety of candidate $U(1)_R$ symmetries, none of which are obviously the superconformal $U(1)_R$. That is, if $R_0$ is some valid $U(1)_R$ symmetry, so too is
\be \label{eqn:rsym}
R_t = R_0 + \sum_i s_i F_i
\ee
where $F_i$ are all the non-R flavor charges of the global symmetry group $\mathcal{F}.$ The superconformal $U(1)_R$ corresponds to some specific choice of the $s_i$. 

In truth, the situation is not quite so dire; the superconformal $U(1)_R$ is expected to commute with non-Abelian flavor symmetries, so we can restrict the linear combinations in (\ref{eqn:rsym}) to only Abelian flavor generators. Moreover, if there is some sort of charge conjugation symmetry, then the $U(1)_R$ should commute with that as well,  leaving only the $F_i$ commuting with charge conjugation. For the simplest example of SQCD, these conditions are sufficient to imply that the superconformal $U(1)_R$ can't mix with any generators of the global symmetries $SU(F) \times SU(F) \times U(1)_B,$ so that the superconformal $U(1)_R$ may be uniquely determined by the vanishing of the ABJ anomaly at the superconformal fixed point. For theories with additional matter content, however, one must somehow account for potential contributions from all possible abelian flavor symmetries.

The solution to this obstruction is a clever procedure called $a$-maximization \cite{Intriligator:2003jj}, which amounts to the observation that the superconformal $R$ charges are those that locally maximize the central charge $a$. Recall that $a$ is the coefficient of the curvature term in the trace of the 4d energy-momentum tensor,
\be
\langle T_\mu^\mu \rangle =  - \frac{a}{16 \pi^2} (\tilde R_{\mu \nu \rho \sigma})^2 + ...
\ee
Conveniently, supersymmetry allows us to compute $a$ for a given theory in terms of traces of R-charges. The result, due to \cite{Anselmi:1997am, Anselmi:1997ys}, is 
\begin{eqnarray}
a = \frac{3}{32} \left[ 3 \Tr R^3 - \Tr R \right],
\end{eqnarray}
where $\Tr R = \sum_i |r_i| (R_i - 1)$ is the sum over fermionic $R$-charges of the matter fields $i$ in the theory, weighted by their dimensions $r_i$. The insight of $a$-maximization is that the correct values of the $s_i$ corresponding to the superconformal $U(1)_R$ charge are given when the trial $a$ function
\be
a_t(s_i) = \frac{3}{32} \left[ 3 \Tr R_t^3 - \Tr R_t \right]
\ee
has a local maximum as a function of the $s_i$. The $R$-charges given by the $a$-maximization procedure are precisely those appearing in the superconformal $U(1)_R$, and hence give the correct scaling dimension of gauge-invariant chiral primary operators in cases where it may not be determined by simpler means.\footnote{Of course, the central charge $a$ is of interest for more than simply $a$-maximization; for some time, this $a$ was conjectured to obey a 4d analogue of the $c$-theorem,  although counterexamples have subsequently been found \cite{Shapere:2008un}.} 

Of course, as with so many clever things, this solution is contingent on being able to identify all the global $U(1)$ symmetries at the infrared fixed point. Clearly, if only a subset of the total $U(1)$ global symmetries have been identified, $a$-maximization over this incomplete subset will generally yield an incorrect result. Although it's sometimes sufficient to identify the $U(1)$ global symmetries in the ultraviolet, it is frequently the case that accidental global $U(1)$ symmetries emerge in the infrared. Most commonly these accidental $U(1)$'s are associated with gauge-invariant operators $\MO$ saturating the unitarity bound in the IR. When such a field goes free, there arises a new $U(1)$ global symmetry associated with rotations of $\MO$. These accidental $U(1)$s will spoil the $a$-maximization procedure unless appropriately accounted for.\footnote{It is amusing to note that the counterexample to the conjectured $a$-theorem exploits precisely this loophole in the $a$-maximization procedure.}

\subsection{Accounting for accidental $U(1)$s}

In the event that an operator $\CO$ hits the unitarity bound, the $a$-maximization procedure will only yield correct $R$-charges provided that $a(R_t)$ is modified to account for $\CO$ going free. In principle, this may be accomplished by replacing the putative contribution from $R_{\CO}$ by the free-field value of $R_{\CO} = 2/3$, via
\be
a(R_t) \to  a(R_t) + a(2/3) - a(\CO) 
\ee
This may be implemented physically using a procedure developed in \cite{Barnes:2004jj}. If $\CO$ transforms as some representation $r_\CO$ of the global symmetry group, consider introducing an additional vector-like pair of gauge-invariant superfields $L, M$ to the theory, where $M$ transforms in the same flavor representation as $\CO$ and $L$ transforms in the appropriate conjugate representation. In addition to the new fields, include also a superpotential
\be
W_{LM} = L(\CO + h M).
\ee
Treating $h$ as a perturbation, when $h=0$ we see that $M$ is a free field and $R(L) = 2 - R(\CO).$ Now turn on a small $h$; if $R(\CO) > 2/3$ (i.e., when $\CO$ is consistent with the unitarity bound), the term $h L M$ is relevant, so $L$ and $M$ become massive and may be integrated out.  In this case, the theory in the IR is identical to the original theory; the contributions of $L$ and $M$ to anomalies and $a$-maximization cancel entirely. However, when $\CO$ violates the unitarity bound, the picture changes significantly. It's still the case that $R(L) = 2 - R(\CO)$, but now $R(\CO) < 2/3$ implies that the coupling $h$ is irrelevant and flows to zero in the IR. In that case, $M$ is a free field with $R(M) = 2/3$, and the contributions of $L$ and $M$ to $a$ no longer cancel. Indeed, adding $L$ and $M$ to the $a$-maximization procedure entails
\begin{eqnarray}
a(R_t) \to a(R_t) + a(M) + a(L) \\
= a(R_t) + a(2/3) - a(\CO) \\
= a(R_t) + \frac{\dim(r_\CO)}{96} (2 - 3 R_\CO)^2(5-3 R_\CO)
\end{eqnarray} 
Hence the addition of $L$ and $M$ to the theory precisely fixes $a$ in the desired fashion when the field $\CO$ goes free. Naturally, this prescription may be generalized to account for any number of operators hitting the unitarity bound.

Although the addition of $L$ and $M$ was introduced as a somewhat {\it ad hoc} procedure for fixing up the $a$-maximization procedure, such fields must additionally be accounted for in all anomaly calculations involving flavor symmetries of the SCFT \cite{Poland:2009yb}. In particular, the effects of $L$ and $M$ must be included in the running of the Standard Model gauge coupling $g_5$ when $SU(5)_{SM}$ is embedded in a weakly gauged subgroup of the superconformal global symmetries. Indeed, the inclusion of these contributions is crucial in correctly determining the effects of the SCFT sector on the running of $g_5$, particularly when determining the scale of potential Standard Model Landau poles.

At first glance, this may seem somewhat unusual; the fields $L$ and $M$ were introduced merely to account for gauge-invariant operators going free in the $a$-maximization procedure. The necessity of accounting for their contributions to other anomalies becomes most transparent when viewed from the perspective of the composite degrees of freedom in the IR. As noted in \S 2, the contribution of SCFT fields to the running of the gauge coupling $g_5$ is equivalent to their contribution to the $U(1)_R SU(5)_{SM}^2$ global anomaly of the SCFT. As such, anomaly-matching guarantees that these contributions must be the same whether computed in terms of the UV or IR degrees of freedom.

Consider then the contribution of a composite operator $\CO$ to the  NSVZ $\beta$-function for $g_5$:
\be
\Delta \mathfrak{b} = (1- \gamma_{\CO}) T(r_\CO) = 3 (1- R_\CO) T(R_\CO)
\ee
When $R_\CO$ violates the unitarity bound, the na\"{i}ve contribution from $\gamma_\CO \neq 0$ computed via $a$-maximization is incorrect. But notice that when $\CO$ goes free, the contributions from the corresponding $L$ and $M$ to the NSVZ $\beta$-function for $g_5$ are given by
\beq
\sum_{i = L,M} (1 - \gamma_i) T(r_i) = \sum_{i = L, M} 3(1 - R_i) T(r_i)  = 3 (1 - 2/3) T(r_M) + 3 (1 - R_L) T(r_L) \\ \nonumber
 = T(r_\CO) + 3(R_\CO -1) T(r_\CO) =  \gamma_\CO T(r_\CO)
 \eeq
which precisely cancels the na\"{i}ve contribution from $\CO$ violating the unitarity bound and enforces $\gamma_\CO = 0.$ Thus incorporating the effects of $L$ and $M$ in the running of $g_5$ does not merely fix the $a$-maximization procedure; it also fixes the contribution of composite fields to all global anomalies of the SCFT (and hence also to $\beta_{g_5}$). 

In general, these additional contributions have the effect of lowering the contribution of the SCFT to $\beta_{g_5}$ (as one might expect, since the na\"{i}ve $\gamma_\CO$ are negative and increase $\mathfrak{b}$). As such, they play a key role in determining what candidate superconformal sectors may explain the flavor hierarchy before generating a Landau pole for the Standard Model gauge coupling.

\section{Simple models with $SU(N)$}

With these tools in hand, let us now turn to a series of simple vector-like models of superconformal flavor whose anomalous dimensions may be calculated using $a$-maximization. We will begin with models where the superconformal sector consists of an $SU(N)$ gauge group, adjoint superfield $A$, and some number of fundamental and antifundamental flavors. In \cite{Poland:2009yb} it was claimed that such models are incapable of generating a sufficient flavor hierarchy before hitting a Landau pole in $g_5$. We will find, to the contrary, that in many cases the Landau poles are sufficiently remote once the contributions from SCFT states to $\beta_{g_5}$ are correctly accounted for.\footnote{Following correspondence with the authors of \cite{Poland:2009yb}, their results have been revised to agree with those found here.}

\subsection{SQCD with an adjoint}

Before focusing on specific models of superconformal flavor, it is worthwhile to review a few useful facts about $\mathcal{N} = 1$ supersymmetric $SU(N)$ QCD with $F$ (anti)fundamental flavors $Q$ ($\tilde Q$) and a single adjoint chiral superfield $A$. The theory with a polynomial superpotential for the adjoint was first studied extensively in \cite{Kutasov:1995ve, Kutasov:1995np, Kutasov:1995ss}, and later re-examined using $a$-maximization \cite{Kutasov:2003iy}. The dynamics of the theory are rendered fairly simple by the addition of a simple superpotential for the adjoint of the form
\be
W = \frac{s_0}{k+1} \Tr A^{k+1}
\ee
Such theories possess and $SU(F) \times SU(F) \times U(1)_B \times U(1)_R$ global symmetry; the transformation properties of $Q, \tilde Q,$ and $A$ under the gauge and global symmetries are shown in  Table~\ref{symmetries}.

\begin{table}[h!]
\caption{Transformation properties of matter fields in SQCD with an adjoint}
\begin{center}
\begin{tabular}{|c|c|c|c|c|c|}
\hline
 & $SU(N)$ & $SU(F)$ & $SU(F)$ & $U(1)_B$ & $U(1)_R$ \\ \hline
 $Q$ & $\square$ & $\square$ & $\mathbf{1}$ & $1$ & $1 - \frac{2}{k+1} \frac{N}{F}$ \\
 $\tilde{Q}$ & $\overline \square$ & $\mathbf{1}$ & $\overline \square$ & $-1$ & $1 - \frac{2}{k+1} \frac{N}{F}$ \\
 $A$ & $\mathbf{Adj}$ & $\mathbf{1}$ & $\mathbf{1}$ & 0 & $\frac{2}{k+1}$ \\ \hline 
\end{tabular}
\end{center}
\label{symmetries}
\end{table}%

In general, it is often interesting to study the theory with a more general polynomial superpotential,
\be
W = \sum_{i = 0}^{k-1} \frac{s_i}{k+1-i} \Tr A^{k+1-i}
\ee
which breaks the remaining $R$-symmetry for nonzero $s_i$. The adjoint superpotential typically breaks the gauge group $SU(N) \to SU(r_1) \times ... \times SU(r_k) \times U(1)^{k-1}$ in the infrared. 

Such theories possess stable vacua provided $F \geq N / k.$ In the far infrared, they may be described in terms of a dual ``magnetic'' supersymmetric gauge theory consisting of a magnetic gauge group $SU(k F - N),$ $F$ magnetic quarks and antiquarks $q, \tilde q$, a magnetic adjoint $a$, and gauge singlets $M_j \sim \tilde Q A^{j-1} Q$ representing mesons of the UV theory. Examination of the beta function for the magnetic gauge coupling reveals that the theory is interacting at its IR fixed point provided  $N < \frac{2 k -1}{2} F.$

\subsection{A ten-centered model \label{SUN-ten}}

Perhaps the simplest vector-like model of superconformal flavor with a rank-two tensor is an $SU(N)$ gauge theory with adjoint $A$ and $F = 10$ fundamental and antifundamental flavors. For simplicity, we will also assume that the term $\Tr A^3$ is marginal at the conformal fixed point. We may embed $SU(5)_{SM}$ in the $SU(F) \times SU(F)$ global symmetry group as shown in Table~\ref{simplest}.

\begin{table}[h]
\caption{Embedding of a ten-centered model with $F=10$}
\begin{center}
\begin{tabular}{|c|c|c|}
\hline
 & $SU(5)_{SM}$ & $SU(N)$ \\ \hline
$Q_1 + \overline Q_2$ & $\mathbf{ 5 + \overline 5}$ & $\square$ \\
 $\overline Q_1 + Q_2$ & $\mathbf{\overline 5 + 5}$ & $\overline \square$ \\
 $A$ & $\mathbf{1}$ & $\mathbf{Adj.}$ \\  \hline
\end{tabular}
\end{center}
\label{simplest}
\end{table}%

The ($SU(N)$) gauge-invariant mesons transform under $SU(5)$ as
\be
(Q_1 + \overline{Q}_2)(\overline{Q}_1 + Q_2) = 2\times \mathbf{24} + 2 \times \mathbf{1} + \mathbf{10} + \mathbf{15} + \mathbf{\overline{10}} + \mathbf{\overline{15}}
\ee
With this field content, the most general couplings involving two SCFT fields and one Standard Model field are those incorporating the $T_1$ and $T_2$ fields.  As such, our marginal superpotential terms at the conformal fixed point are
\beq
W = T_1 \overline Q_1 \overline Q_2 + T_2 \overline Q_1 A \overline Q_2 + A^3
\eeq
For the theory with $k = 2$ and $F = 10,$ we require $N < 15$ for the theory to be interacting at the fixed point and $N < 20$ to have stable vacua. We will also find $R_{T_2} > 2/3$ only for $N >10$, which gives us a window $10 < N < 15$ for this particular theory. By assumption, the $\beta$ function for the $SU(N)$ gauge coupling vanishes at the fixed point, and the above operators are held to be marginal. The $R$ charges, and hence the scaling dimensions, of the Standard Model fields $T_1, T_2$ may then be computed via $a$-maximization, the numerical results of which are reserved for Table~\ref{firstmodel} in Appendix A. We find that several mesons are free fields at the fixed point: $Q_1 Q_2$, $Q_1 \overline Q_1$, $Q_2 \overline Q_2$, and $\overline Q_1 \overline Q_2$. Of these, only the $\overline{\mathbf{15}}$ of the last meson needs to be accounted for in the $a$-maximization procedure, since the $\overline{\mathbf{10}}$ component is set to zero in the chiral ring due to the superpotential couplings with $T_1$. 

 For $N = 11, 12$, the R-charges of SM fields are too small to generate the observed flavor hierarchy over any range of running. However, the theories with $N = 13, 14$ work beautifully. In both cases, a sufficient flavor hierarchy may be generated before $g_5$ hits a Landau pole. In order for this to work, it is crucial to correctly account for the effects of the mesons $Q_1 Q_2$, $Q_1 \overline Q_1$, $Q_2 \overline Q_2$, and $\overline Q_1 \overline Q_2$ going free when computing their contribution to $\beta(g_5)$. Thus we find that $SU(N)$ SQCD with an adjoint and $F = 10$ fundamental flavors provides a suitable model of superconformal flavor.

It is tempting to consider the same theory with marginal operator $\Tr A^4$ at the conformal fixed point. For such a theory, we require $N < 25$ in order to be interacting and $N < 30$ for stability. We also find that $T_2$ violates the unitarity bound for $N < 12$, so we are interested in values $11 < N < 25$.  The constraints on $R$-charges and results of $a$-maximization are reserved for Table~\ref{secondmodel}. For sufficiently large $N$ -- specifically, for $21 \leq N \leq 24$ -- a sufficient flavor hierarchy may be generated before $g_5$ hits a Landau pole.

\subsection{A more democratic model}

Although ten-centered models capture much of the essential features of the flavor hierarchy, it is worth exploring whether a more complete hierarchy may be generated by coupling the SCFT to both $T_i$ and $\bar F_i$ fields of the Standard Model. Extending our $SU(N)$ model to accommodate couplings to additional Standard Model representations is fairly simple; it requires only enlarging the flavor symmetry to $F>10$. The simplest such model involves $F =11$ fundamental and antifundamental flavors of the SCFT. The matter content and transformation properties under $SU(5)_{SM}$ are shown in Table~\ref{Nf11matter}.

\begin{table}[h] \label{Nf11matter}
\caption{Matter content for $SU(N)$ theory with $F = 11$}
\begin{center}
\begin{tabular}{|c|c|c|}
\hline
 & $SU(5)_{SM}$ & $SU(N)$ \\ \hline
$Q_1 + \overline Q_2 + Q_0$ & $\mathbf{ 5 + \overline 5 + 1}$ & $\square$ \\
 $\overline Q_1 + Q_2 + \overline Q_0$ & $\mathbf{\overline 5 + 5 + 1}$ & $\overline \square$ \\
 $A$ & $\mathbf{1}$ & Adj. \\  \hline
\end{tabular}
\end{center}
\label{default}
\end{table}

Naturally, there is a significant increase in the number of $SU(N)$ gauge invariants transforming nontrivially under $SU(5)$:
\be
(Q_1 + \overline Q_2 + Q_0)(\overline Q_1 + Q_2 + \overline Q_0) = 2\times \mathbf{24} + 3 \times \mathbf{1} + \mathbf{10} + \mathbf{15} + \mathbf{\overline{10}} + \mathbf{\overline{15}} + 2 \times \mathbf{5} + 2 \times \mathbf{\overline{5}}
\ee

For clarity, the transformation properties of the $SU(N)$ gauge-invariant mesons under $SU(5)$ is shown in detail in Table~\ref{mesondecomp}.

\begin{table}[h]
\caption{Meson decomposition under $SU(5)$}
\begin{center}
\begin{tabular}{|c|c||c|c|}
\hline
Meson & $SU(5)$ & Meson & $SU(5)$ \\ \hline
$Q_1 Q_2$ & $\mathbf{10 + 15}$ & $Q_1 \overline Q_0$ & $\mathbf{5}$ \\ 
$Q_1 \overline Q_1$ & $\mathbf{24 +1}$ & $Q_2 Q_0$ & $\mathbf{5}$ \\
$Q_2 \overline Q_2$ & $\mathbf{24+1}$ & $\overline Q_1 Q_0$ & $\mathbf{\overline{5}}$ \\
$\overline Q_1 \overline Q_2$ & $\mathbf{\overline{10} + \overline{15}}$ & $\overline Q_2 \overline Q_0$ & $\mathbf{\overline{5}}$ \\
 & & $Q_0 \overline{Q}_0$ & $\mathbf{1}$ \\ \hline 
\end{tabular}
\end{center}
\label{mesondecomp}
\end{table}

Many of these mesons go free at the conformal fixed point: $Q_1 Q_2$, $Q_1 \overline Q_1$, $Q_2 \overline Q_2,\overline Q_1 \overline Q_2, \overline Q_1 Q_0, \overline Q_2 \overline Q_0,$ and $Q_0 \overline Q_0$. The $\mathbf{\overline{10}}$ component of $\overline Q_1 \overline Q_2$ will be set to zero in the chiral ring, as will the $\mathbf{5}$s that couple to Standard Model matter. As for the vacua of the theory and the range of parameters, with our customary $\Tr A^3$ deformation we require $N < 17$ for the fixed point to be interacting and $N \leq 22$ for stability of the vacuum.

Our marginal couplings at the fixed point now are\footnote{Of course, it is also now possible to couple the SCFT fields to the Higgses $H_u$ and $H_d$; it is technically natural to turn these couplings off, which we will do here for simplicity. For a discussion of the potential complications that arise from coupling SCFT fields to $H_u$ and $H_d$, see  \cite{Poland:2009yb}.} 
\beq
W = T_1 \overline Q_1 \overline Q_2 + T_2 \overline Q_1 A \overline Q_2 + \overline F_1 (Q_1 \overline Q_0 + Q_2 Q_0)+ \overline F_2 (Q_1 A \overline Q_0 + Q_2 A Q_0) + A^3
\eeq
The results of the $a$-maximization procedure are shown in Table~\ref{thirdmodel}. We see that at $N = 11$ all the mesons of the SCFT are exactly free, and violate the unitarity bound for $N > 11$. So we can use our usual techniques to analyze the theory in the window $10 < N < 17$. The results are fairly encouraging; for $N > 13$ it is possible to generate a sufficient hierarchy for both the $T_i$ and $\overline F_i$ before hitting a Landau pole of $SU(5)$. 

\subsection{Coupling to the adjoint Higgs}

One way to address the potential Landau pole in the $SU(5)$ gauge coupling is to find other ways to reduce naive contributions from Standard Model GUT fields. Large anomalous dimensions do precisely that; since the contribution to $\beta_{g_5}$ of a matter field in the representation $r_i$ is proportional to $T(r_i) (1-\gamma_i)$, it's clear that large, positive anomalous dimensions $\gamma_i$ can slow somewhat the progression of $g_5$ towards its Landau pole.

A simple way to implement this idea is to couple the superconformal sector to the $SU(5)_{SM}$ adjoint Higgs field $\Sigma$ responsible for breaking $SU(5) \to SU(3) \times SU(2) \times U(1).$ Such couplings are, in general, allowed by the symmetries of models considered here, and are not unreasonable to include among the marginal interactions at the superconformal fixed point.
 
Consider, e.g., the model of \S~\ref{SUN-ten}, incorporating now additional couplings to the adjoint Higgs $\Sigma$. The allowed interactions are now

\beq
W_{int} = T_1 \overline Q_1 \overline Q_2 + T_2 \overline Q_1 A \overline Q_2 + (Q_1 \overline Q_1 + Q_2 \overline Q_2) \Sigma + A^3
\eeq

In principle this gives two extra constraints and one extra unknown, but in fact the two new terms are identical equations, so we still need $a$-maximization to solve for the R-charges. Doing $a$-maximization on the $A^3$ theory gives us only changes to the value of $\mathfrak{b}$; this dramatically improves the window of running for SU(5) couplings while preserving the nice predictions of the undeformed theory. There is a small trade-off in that the $\mathbf{24}$ component of the linear combination $Q_1 \overline Q_1 + Q_2 \overline Q_2$ is now set to zero in the chiral ring, but nonetheless the net effect is to lower $\mathfrak{b}$ significantly and thus render the Landau pole more remote.\\

\section{Simple models with $Sp(2N)$}

Although it is compelling that something as simple as $SU(N)$ SQCD with an adjoint leads to suitable models of superconformal flavor, it's useful to consider related models with different gauge groups. Symplectic groups, in particular, offer more ``compact'' theories of flavor, in the sense that Standard Model $SU(5)$ may be more efficiently embedded in their flavor symmetries. 

In this section we will focus on $\mathcal{N} = 1$ supersymmetric $Sp(2N)$ gauge theory\footnote{Here we are choosing notation such that $Sp(2) \sim SU(2)$} with $2F$ fundamental flavors $Q$ and an antisymmetric tensor $A$.\footnote{The related $Sp(2N)$ theory with {\it symmetric} tensor $A$, studied extensively in \cite{Leigh:1995qp}, is less suitable for these simple models of flavor due to the different symmetry properties of the mesons $QQ$ and $QAQ$.}  The IR behavior of $Sp(2N)$ theories with an antisymmetric tensor and polynomial superpotential $\Tr A^{k+1}$ was studied in detail in \cite{Intriligator:1995ff}, while the theory without polynomial superpotential was analyzed using $a$-maximization in \cite{Poland:2009px}. With some malice aforethought, we will focus here on the $k=2$ superpotential with marginal operator $\Tr A^3$. This theory is interacting in the IR provided $N < \left( k - \half \right) F - 2(k-1)$ and possesses stable vacua provided $N < k F$. The transformation properties of the matter fields under the relevant gauge and global symmetries is shown below in Table~\ref{spsymmetries}.

\begin{table}[h!]
\caption{Transformation properties of matter fields in $Sp(2N)$ with an antisymmetric tensor}
\begin{center}
\begin{tabular}{|c|c|c|c|c|c|}
\hline
 & $Sp(2N)$ & $SU(2F)$ & $U(1)_R$ \\ \hline
 $Q$ & $\square$ & $\square$ & $1 - \frac{2(N+k)}{(k+1)F}$ \\
 $A$ & $\mathbf{Anti.}$ & $\mathbf{1}$ & $\frac{2}{k+1}$ \\ \hline 
\end{tabular}
\end{center}
\label{spsymmetries}
\end{table}%

\subsection{A ten-centered model}

As a warmup, let us begin with the simplest $Sp(2N)$ theory: $Sp(2N)$ with $2F = 10$ flavors of fundamental quark $Q$ and antisymmetric tensor $A$. This theory was treated in \cite{Poland:2009yb}; we review their results here before moving on to a more general model with larger flavor symmetry. The embedding is shown in Table~\ref{spsimplest}.

\begin{table}[h!]
\caption{Embedding of a ten-centered $Sp(2N)$ model with $2F=10$}
\begin{center}
\begin{tabular}{|c|c|c|}
\hline
 & $SU(5)_{SM}$ & $Sp(2N)$ \\ \hline
$Q + \overline Q$ & $\mathbf{ 5 + \overline 5}$ & $\square$ \\
 $A$ & $\mathbf{1}$ & $\mathbf{Anti.}$ \\  \hline
\end{tabular}
\end{center}
\label{spsimplest}
\end{table}%

The mesons of the SCFT then transform under $SU(5)_{SM}$ as
\beq
(Q + \overline Q) J (Q + \overline Q) = \mathbf{24} +  \mathbf{1} + \mathbf{10 + \overline{10} }
\eeq

As in the $SU(N)$ theory with $F=10$, there are no $\mathbf{5}$ representations to combine with the $\overline F_i$ of the SM, making this a purely ten-centered model. In this case our desired couplings to SM fields are (including the customary cubic superpotential for the antisymmetric tensor)
\beq
W = T_1 \overline Q \overline Q + T_2 \overline Q A \overline Q + \Tr A^3
\eeq
For $k = 2$ and $F= 5$ we require 
$N\leq 5$ in order for the theory to possess an interacting fixed point. We find that the gauge invariant chiral operators $Q Q$ and $Q \overline Q$ go free in the range of interest, while $\overline Q \overline Q$ and $\overline Q A \overline Q$ are set to zero in the chiral ring. There are no baryons in the chiral ring of this theory, since putative baryons of an $Sp(2N)$ gauge theory may be expressed in terms of mesons. 

The constraints and $R$-charges computed via $a$-maximization are reserved for Table~\ref{sp2n-ten} of the the appendix. Given the constraint on $N$, the possible theories are fairly proscribed. However, for $N = 5$ the theory generates a sufficient flavor hierarchy over a small range of energies. Equally attractive is the remoteness of Landau poles; the relative smallness of the additional Standard Model representations introduced by the SCFT ensures that $g_5$ remains perturbative many orders of magnitude above the GUT scale. 

It is fairly straightforward to compute the $R$-charges for the simple extension to the $k = 3$ theory with $2F = 10$ flavors. The virtue of such theories is a larger window of $N$ for which the IR fixed point is interacting -- in this case, for $N \leq 8$. The lowered $R$-charge of $A$ allows the mesons $Q A Q$ and $Q A \overline Q$ to saturate the unitarity bound as well. For $8 \geq N \geq 5$ the outcome is encouraging: adequate flavor hierarchy with Landau poles far from the GUT scale.

\subsection{A more democratic model}

As before, we can consider extending the ten-centered model in \S 6.1 by enlarging the flavor symmetry of the superconformal sector. In this case, the simplest generalization is to increase the number of fundamental flavors to $2F =12$ (recalling that we need an even number of flavors to cancel the global anomaly). As always, we may then weakly gauge an $SU(5)$ subgroup of the flavor symmetry. The corresponding transformation properties of the SCFT fields are shown in Table~\ref{spgeneral}.

\begin{table}[h]
\caption{$SU(5)$ embedding of a democratic $Sp(2N)$ model with $2F=12$}
\begin{center}
\begin{tabular}{|c|c|c|}
\hline
 & $SU(5)_{SM}$ & $Sp(2N)$ \\ \hline
$Q + \overline Q + Q_0 + \overline Q_0$ & $\mathbf{ 5 + \overline 5 + 1 + 1}$ & $\square$ \\
 $A$ & $\mathbf{1}$ & $\mathbf{Anti.}$ \\  \hline
\end{tabular}
\end{center}
\label{spgeneral}
\end{table}%

The enlarged flavor symmetry leads to a plethora of $Sp(2N)$ gauge-invariant chiral operators transforming under $SU(5)_{SM}$, which we list for convenience in Table~\ref{spmesondecomp}. 

\begin{table}[h]
\caption{Meson decomposition of under $SU(5)$}
\begin{center}
\begin{tabular}{|c|c||c|c|}
\hline
Meson & $SU(5)$ & Meson & $SU(5)$ \\ \hline
$Q Q $ & $\mathbf{10}$ & $\overline Q \overline Q$ & $\mathbf{\overline{10}}$ \\ 
$Q \overline Q$ & $\mathbf{24 +1}$ & $\overline Q Q_0$ & $\mathbf{\overline 5}$ \\
$Q Q_0$ & $\mathbf{5}$ & $\overline Q \overline Q_0$ & $\mathbf{\overline{5}}$ \\
$Q \overline Q_0$ & $\mathbf{5}$ & $Q_0 \overline Q_0$ & $\mathbf{1}$ \\ \hline 
\end{tabular}
\end{center}
\label{spmesondecomp}
\end{table}

Assuming our customary cubic superpotential term for the antisymmetric field, the theory possesses stable vacua provided $N < 12$ and is at an interacting IR fixed point provided $N < 7.$

The candidate marginal couplings at the conformal fixed point are thus
\beq
W = T_1 \overline Q \overline Q + T_2 \overline Q A \overline Q + F_1 (Q Q_0 + Q \overline Q_0) + F_2 (Q A Q_0 + Q A \overline Q_0) + \Tr A^3
\eeq
As always, the gauge-invariant chiral operators with marginal couplings to Standard Model states are set to zero in the chiral ring ($\overline Q \overline Q, \overline Q A \overline Q,$ and the linear combinations $Q Q_0 + Q \overline Q_0, Q A Q_0 + Q A \overline Q_0$). Of the remaining chiral operators, $QQ, Q \overline Q, \overline Q Q_0, \overline Q \overline Q_0,$ and $Q_0 \overline Q_0$  saturate the unitarity bound and must be accounted for accordingly in the $a$-maximization procedure.

The superconformal $R$-charge assignments for this theory are shown in Table~\ref{spgenmodel}. For $N =6$ the theory produces a sufficient flavor hierarchy for both the $T_i$ and $\overline F_i$ well below any potential Landau poles in $g_5$. The $k = 3$ theory with $2F = 12$ is essentially identical in features, albeit with a much larger window of colors (ranging up to $ N < 11$ for an interacting fixed point).

\section{Discussion}

Thus far we have seen that a variety of models based on $SU(N)$ and $Sp(2N)$ superconformal gauge theories with rank-two tensor fields may give rise to the Standard Model flavor hierarchy above the GUT scale. However, our treatment has elided a few significant details that warrant some consideration -- in particular, the effect of Standard Model field couplings on the superconformal fixed point, as well as the details of conformal symmetry breaking and decoupling -- to which we now turn.

\subsection{Saturating the unitarity bound}

In the preceding sections, we have been interested in superpotential interactions coupling Standard Model and SCFT fields of the form $\delta W = \Phi_i \MO$, where $\MO$ is a gauge-invariant chiral operator comprised of matter fields of the SCFT. Thus far we have treated such interactions as a small deformation away from the original superconformal fixed point of the SCFT sector, but it is worth examining whether this approximation is completely justified. It is often the case in the undeformed SCFT that the scaling dimension of $\MO$ saturates the unitarity bound, at which point an accidental $U(1)$ symmetry emerges to enforce $R_\MO = 2/3.$ When the SCFT has a dual description in which the magnetic dual of $\MO$ is a free field, we generally interpret saturation of the unitarity bound as an indication that the field $\MO$ has gone free.  

The issue becomes somewhat more convoluted in the models considered here, where $\MO$ is coupled additionally to Standard Model fields $\Phi_i$ by marginal superpotential interactions. In that case, when $\MO$ hits the unitarity bound it is no longer the case that an accidental $U(1)$ emerges to enforce $R_\MO = 2/3$, but rather $R_\MO < 2/3$ is allowed. This $R$ charge is not in conflict with the unitarity bound, as the $F$ term for $\Phi_i$ sets $\MO$ to zero in the chiral ring, so that the unitarity bound no longer pertains. One might become concerned about whether the interaction $\Phi_i \MO$ in this case truly amounts to a small deformation of the superconformal fixed point, since it involves positing a marginal interaction between a Standard Model field and an otherwise-free operator.\footnote{We thank Dan Green for an extensive discussion of this point. For further discussion, see also \cite{Baumann:2010ys}.}

 This question becomes fairly central in the models considered above, where generating an adequate flavor heirarchy before hitting a Standard Model Landau pole requires $R(\MO) < 2/3$ (for at least one such $\MO$) in every case.\footnote{Indeed, the only viable vector-like model of superconformal flavor that does not require $R(\MO) < 2/3$ for some $\MO$ is the $Sp(2N)$ theory with no polynomial superpotential for $A$, studied in \cite{Poland:2009yb}. However, this theory requires significantly more decades of running above the decoupling scale $\Lambda_*$, and may not fit between $M_{GUT}$ and $M_{Pl}$.} 

Thankfully, the new fixed points reached by coupling Standard Model fields to the SCFT are fairly well understood. Turning off the Standard Model gauge coupling and Yukawa interactions reduces these models to variations on $SU(N)$ SSQCD (analyzed via $a$-maximization in \cite{Barnes:2004jj}) and its $Sp(2N)$ generalization. In this case, the role of the gauge singlets of SSQCD is played by Standard Model superfields. Although our models also differ from SSQCD by the inclusion of rank-two tensor fields, these do not significantly modify the relevant details. To understand the fixed point in detail, let us review the results of \cite{Barnes:2004jj}. Consider $SU(N)$ SQCD with $F$ flavors $Q_i, \tilde Q_i$ and $F'$ additional flavors $Q_{i'}', \tilde Q_{i'}'$, as well as $F'^2$ singlets $S^{i' j'}$ with superpotential coupling
\beq
W = h S^{i' j'} Q_{i'}' \tilde Q_{j'}'
\label{SSQCDint}
\eeq
For $h=0$, the IR fixed point is simply that of SQCD with $F + F'$ flavors, which we know to have an interacting fixed point for $3 N / 2 < F + F' < 3 N$. Turning on $h \to h_* \neq 0$ amounts to a relevant deformation driving the theory to a new family of SCFTs in the IR, of which the usual fixed points of SQCD are special cases.

	The fixed point may also be described by a dual $SU(F + F' - N)$ gauge theory. In the dual theory, the interaction (\ref{SSQCDint}) corresponds to a mass term for the singlets $S$ and the mesons $M' \sim Q' \tilde Q'$, which may be integrated out. The remaining matter content at the fixed point consists of $F$ flavors of magnetic quarks $q', \tilde q'$; $F'$ flavors $q, \tilde q$; an $SU(F) \times SU(F)$ bifundamental meson $M_{ij}$, and $SU(F) \times SU(F')$ bifundamental mesons $P_{ij'}, P'_{ij'}$ with superpotential interaction
\beq
W = M q' \tilde q' + P q' \tilde q + P' \tilde q' q
\eeq
The duality map for various gauge-invariant operators is 
\beq
Q \tilde Q \to M, \; \; \; \; S \to - q \tilde q, \; \; \; \; Q \tilde Q' \to P, \; \; \; \; Q' \tilde Q \to P', \; \; \; \; Q^r Q'^{N - r} \to q'^{F-r} q^{F' - N + r}
\eeq
 Significantly, although $M'$ and $S$ have been integrated out of the dual theory, there remains a gauge-invariant chiral operator (identified with $-q \tilde q$) that has the same quantum numbers as the original singlets $S$. 

Both the original theory and its dual share a $SU(F)_L \times SU(F)_R \times SU(F')_L \times SU(F')_R \times U(1)_B \times U(1)_{B'} \times U(1)_F \times U(1)_{R_0}$ flavor symmetry; the axial $SU(F + F')$ flavor symmetry is broken to $SU(F) \times SU(F') \times U(1)_F$ by $h \neq 0$. The $R$ charges may be determined in the original theory by carrying out the $a$-maximization procedure subject to the constraints 
\beq
N + F (R(Q) - 1) + F' (R(Q') - 1) = 0, \; \; \; \; \; \; \; R(S) + 2 R(Q') = 2
\eeq
It's clear that the $R$ charge of $Q$ will differ from that of $Q'$; this is because the $R$-symmetry can mix with the $U(1)_F$ flavor symmetry, under which $Q, Q'$ have opposite charges. Significantly, this implies in the original theory that $R(S) = 2 - 2 R(Q')$, irrespective of whether $Q' \tilde Q'$ violates the unitarity bound. The duality map relates these $R$ charges to those of the dual theory, such that
\beq
2 R(Q) = R(M), \; \; \; \; R(S) = 2 R(q), \; \; \; \; R(Q) + R(Q') = R(P)
\eeq
In this case, the duality map implies $2 R(q) = 2 - 2 R(Q')$, so that the gauge invariant operator $q \tilde q$ inherits the $R$ charge and scaling dimension of the singlets $S$. 

Having established the duality map, it is fairly straightforward to understand the results of $a$-maximization in both the original theory and its dual. For fixed $F' / F,$ as $N/F$ is increased the theory goes successively through the phases: free electric fixed point; interacting fixed point with no mesons free; interacting fixed point with only $M = Q \tilde Q$ free; free magnetic fixed point. The meson $M'$ does not appear in the phase diagram, as it has been set to zero in the chiral ring by $F_S$ in the original theory, and equivalently has been integrated out in the dual theory. Nonetheless, in the original theory the field $S$ gains a large anomalous dimension from its coupling to $M' = Q' \tilde Q'$, while in the dual theory the same anomalous dimension is developed by the dual gauge-invariant operator $- q \tilde q$. These results hold whether or not $M' = Q' \tilde Q'$ appears to violate the unitarity bound.

In terms of the models considered in Sections 4 and 5, the results are entirely analogous, although the Standard Model fields $\Phi_i$ transform under different representations of the flavor symmetry than the flavor bifundamental $S$ of SSQCD. In terms of the original variables, an interaction $\Phi_i \MO$ sets $\MO$ to zero in the chiral ring and fixes $R(\Phi_i) = 2 - R(\MO)$ irrespective of whether $\MO$ violates the unitarity bound. In terms of dual variables, $\Phi_i$ and $\MO$ are integrated out, but there is a gauge invariant chiral operator $\tilde \Phi_i$ with the same quantum numbers and $R$ charge as the original $\Phi_i$. In this case, the low energy Standard Model degrees of freedom may be thought of as composites of the dual gauge group. Thus there appears to be nothing inconsistent about generating the flavor hierarchy by coupling Standard Model fields to SCFT operators that go free at the undeformed superconformal fixed point.

\subsection{Decoupling}

Thus far we have remained agnostic about what happens at and below the scale $\Lambda_*$ at which conformality is broken and the theory flows away from its conformal fixed point. It certainly is necessary to decouple the SCFT fields carrying Standard Model charges, lest they come into conflict with observational limits on charged exotics. Thankfully, this may be accomplished easily in vector-like models simply by giving a vector mass to the fundamental quarks and antiquarks of the SCFT sector. 

There are a variety of controlled ways of breaking conformal invariance. Perhaps the most typical way involves turning on vector masses for some or all flavors of fundamental matter at a scale $m_Q \sim \Lambda_*$, so that the theory no longer has enough flavors to remain conformal. Determining the correct IR degrees of freedom after conformal symmetry breaking is a fairly delicate matter; for a detailed discussion, see Appendix B of  \cite{Poland:2009yb}. 

An alternative is to include a mass $m_A \sim \Lambda_*$ for the rank-two tensor $A$, along the lines of \cite{Schmaltz:2006qs}. Below the scale $m_A$, $A$ may be integrated out, leaving SQCD (or possibly a product group of SQCD theories, in the event that the vev of $A$ breaks the original gauge group) with too few flavors to remain conformal. Thus the theory flows to a free fixed point that may be described in terms of the dual SQCD degrees of freedom. Breaking conformality in this fashion raises the possibility that the remaining degrees of freedom may break supersymmetry as in \cite{Intriligator:2006dd, Amariti:2010sz}, although remaining matter charged under the Standard Model must still be decoupled in a controlled fashion.

\section{Conclusion}

The pattern of Standard Model flavor poses a considerable puzzle to theoretical physics; both the replication and hierarchy of fermion masses are without obvious explanation. It is exciting that a superconformal sector coupled to the Standard Model may generate the observed fermion mass hierarchy from complete flavor anarchy over just a few decades in energy. Such a scenario, moreover, may not be entirely fantastic; many ultraviolet completions of the Standard Model give rise to additional gauge groups with bifundamental matter at high energies. If vector-like, these sectors may gracefully decouple at low energies and remain consistent with observational constraints. 

When supersymmetric, these sectors have the virtue of calculability thanks to superconformal symmetry and the $a$-maximization procedure. As such, we may subject them to straightforward tests of consistency. 
Here we have found that simple theories of both $SU(N)$ and $Sp(2N)$ with fundamental matter and a rank-2 tensor field are capable of producing the observed flavor hierarchy before the unified Standard Model gauge coupling hits a Landau pole. Using these results, we have constructed both ten-centered and democratic models of superconformal flavor. It seems that a variety of potential models are viable, over a full range of $\tan \beta$. The challenge now rests in determining which, if any, such models may be realized in nature. Although models operating above the GUT scale are advantageous from the perspective of proton decay and other potentially dangerous baryon number violation, they are generally too remote to yield distinct experimental signatures beyond the observed Yukawa textures. It would be amusing to see if such models may be lowered to accessible energies without running afoul of observational bounds. 

Much progress has been made in recent years towards understanding calculable supersymmetry breaking in vector-like gauge theories, beginning with \cite{Intriligator:2006dd}. Supersymmetry breaking vacua have been found in $SU(N)$ theories with fundamental and adjoint matter \cite{Amariti:2006vk, Craig:2009hf}, making it natural to consider whether both superconformal flavor and supersymmetry breaking may emerge from the same dynamics. The resulting correlations between the patterns of fermion and sfermion flavor may hold the key to explaining the small amount of observed flavor violation, as well as provide indications of the superconformal dynamics in the far ultraviolet. \\

\noindent {\bf Note added upon completion:} After this work was completed, correspondence with the authors of \cite{Poland:2009yb} revealed that the discrepancy in results regarding the viability of $SU(N)$ theories with an adjoint arose from incorrect values of $\mathfrak{b}$ in the original version of  \cite{Poland:2009yb}. Their values and conclusions have subsequently been revised and found to agree with those appearing in Section 4.

\section*{Acknowledgements}

I am grateful to Savas Dimopoulos and John March-Russell for stimulating my interest in theories of flavor anarchy. I would like to thank David Poland and David Simmons-Duffin for extensive discussion of their work; Dan Green for healthy skepticism and comments on the manuscript; and Rouven Essig, Sebastian Franco, Dan Green,  Shamit Kachru, Ann Nelson, David Poland, David Simmons-Duffin, and Gonzalo Torroba for useful conversations on superconformal flavor, $a$-maximization, and related topics. This research is supported by the NSF GRFP and the Stanford Institute for Theoretical Physics under NSF Grant 0756174.

\appendix

\section{Bestiary of R charges}

Here we present the results of the $a$-maximization procedure applied to the various models considered earlier.  The constraints relating various $R$ charges arise from (a) the posited marginal interactions contained in the superpotential, and (b) the vanishing of the ABJ anomaly for the superconformal $U(1)_R$, which corresponds to the vanishing of $\beta$ at the superconformal fixed point. 

We subject the models under consideration to a fairly simple criterion: that they generate an adequate flavor hierarchy over the range $\Lambda_* < \mu < \Lambda$ smaller than the hierarchy $\Lambda_* < \mu < \Lambda_L$ between decoupling and the Landau pole for $g_5$. For all models we compute $\Lambda_L / \Lambda_*$ assuming Standard Model field content, an additional $SU(5)_{SM}$ adjoint Higgs $\Sigma$, and the field content of the superconformal sector. For ten-centered models, we compare this to the ratio $\Lambda / \Lambda_*$ required to get within a factor of $3$ of the observed hierarchy in up-type quark masses. For democratic models, we compare this to both the ratio $\Lambda_T / \Lambda_*$ required to get within a factor of 3 of the up-type quark hierarchy, and the ratio  $\Lambda_F / \Lambda_*$ required to get within a factor of 3 of the lepton mass hierarchy for $\tan \beta = 10$.

\subsection{$SU(N)$ with $F=10$ and $A^3$ superpotential} 

The $R$ charges for this theory are constrained by the posited marginal operators and anomalies to obey
\begin{eqnarray}
2 &=& R_{T_1} + R_{\overline Q_1} + R_{\overline Q_2}   \\ \nonumber
2 &=& R_{T_2} + R_{\overline Q_1} + R_{\overline Q_2} + R_A \\ \nonumber
2 &=& 3 R_A \\ \nonumber
0 &=& N + \frac{5}{2} (R_{Q_1} - 1) + \frac{5}{2} (R_{Q_2} - 1) + \frac{5}{2} (R_{\overline Q_1} - 1) + \frac{5}{2} (R_{\overline Q_2} - 1) + N ( R_A - 1) 
\end{eqnarray}
In the window of interest several mesons go free: $Q_1 Q_2$, $Q_1 \overline Q_1$, $Q_2 \overline Q_2$, and $\overline Q_1 \overline Q_2$. All mesons involving $A$ and all baryons are far from the unitarity bound. Of the free fields, only the $\overline{\mathbf{15}}$ of the $Q_2 \overline Q_2$ needs to be accounted for, since the $\overline{\mathbf{10}}$ part is set to zero in the chiral ring. The $a$-maximization procedure gives us the following charges:

\begin{table}[h]
\caption{R charges for $SU(N)$ theory with $F = 10$ flavors and cubic adjoint superpotential}
\begin{center}
\begin{tabular}{|c|c|c|c|c|c||c|c|c|}
\hline
$N$ & $R_{T_1}$ & $R_{T_2}$ & $R_{Q_{1,2}}$ & $R_{\overline Q_{1,2}}$ & $R_A$  & $\mathfrak{b}$ & $\frac{\Lambda_{L}}{ \Lambda_*}$ & $\frac{\Lambda}{ \Lambda_*}$ \\ \hline
11 & 1.448 & 0.781 & 0.257 & 0.276 & 0.667 & 33.884 & $10^{1.78}$ & -  \\  
12 & 1.572 & 0.905 & 0.186 & 0.214 & 0.667 & 34.528 & $10^{1.75}$ & - \\
13 & 1.705 & 1.039 & 0.119 & 0.147 & 0.667 & 35.925 & $10^{1.68}$ & $10^{1.65}$ \\
14 &  1.849 & 1.182 & 0.058 & 0.076 & 0.667 & 38.080 & $10^{1.58}$ & $10^{1.19}$ \\ \hline
\end{tabular}
\end{center}
\label{firstmodel}
\end{table}%

\subsection{$SU(N)$ with $F=10$ and $A^4$ superpotential}

This theory is a fairly trivial variation of the theory in \S A.1; the $R$ charges are constrained to obey
\beq
2 &=& R_{T_1} + R_{\overline Q_1} + R_{\overline Q_2}  \\ \nonumber
2 &=& R_{T_2} + R_{\overline Q_1} + R_{\overline Q_2} + R_A \\ \nonumber
2 &=& 4 R_A \\ \nonumber
0 &=& N + \frac{5}{2} (R_{Q_1} - 1) + \frac{5}{2} (R_{Q_2} - 1) + \frac{5}{2} (R_{\overline Q_1} - 1) + \frac{5}{2} (R_{\overline Q_2} - 1) + N ( R_A - 1) 
\eeq

In the window of interest several mesons go free: $Q_1 Q_2$, $Q_1 \overline Q_1$, $Q_2 \overline Q_2$, and $\overline Q_1 \overline Q_2$. For $N \geq 19$, we also add to the tally $Q_1 A Q_2$, $Q_1 A \overline Q_1$, $Q_2 A \overline Q_2$, and $\overline Q_1 A \overline Q_2$. At no point do any of the baryons go free.  Of the free fields, only the $\overline{\mathbf{15}}$ of the $Q_2 \overline Q_2$ and $Q_2 A \overline Q_2$ need to be accounted for, since the $\overline{\mathbf{10}}$ parts are set to zero in the chiral ring. The $a$-maximization procedure gives us the following charges:

\begin{table}[h]
\caption{R charges for $SU(N)$ theory with $F =10$ flavors and quartic adjoint superpotential}
\begin{center}
\begin{tabular}{|c|c|c|c|c|c||c|c|c|}
\hline
$N$ & $R_{T_1}$ & $R_{T_2}$ & $R_{Q_{1,2}}$ & $R_{\overline Q_{1,2}}$ & $R_A$  & $\mathfrak{b}$ & $\frac{\Lambda_{L}}{ \Lambda_*}$ & $\frac{\Lambda}{ \Lambda_*}$ \\ \hline
14 & 1.383 & 0.883 & 0.292 & 0.308 & 0.5 & 47.826 & $10^{1.26}$ & - \\
15 & 1.482 & 0.982 & 0.241 & 0.259 & 0.5 & 50.080 & $10^{1.20}$ & $10^{1.95}$ \\
16 & 1.585 & 1.085 & 0.192 & 0.208 & 0.5 & 52.920 & $10^{1.13}$ & $10^{1.49}$ \\
17 & 1.690 & 1.190 & 0.145 & 0.155 & 0.5 & 56.345 & $10^{1.07}$ & $10^{1.33}$ \\
18 & 1.797 & 1.297 & 0.099 & 0.102 & 0.5 & 60.363 & $10^{1.00}$ & $10^{1.21}$ \\
19 & 1.900 & 1.400 & 0.050 & 0.050 & 0.5 & 61.300 & $10^{0.98}$ & $10^{1.11}$ \\ 
20 & 2.000 & 1.500 & 0.000 & 0.000 & 0.5 & 61.000 & $10^{0.99}$ & $10^{1.02}$ \\
21 & 2.100 & 1.600 &-0.050&-0.050 & 0.5 & 61.300 & $10^{0.98}$ & $10^{0.95}$ \\
22 & 2.200 & 1.700 &-0.100&-0.100 & 0.5 & 62.200 & $10^{0.97}$ & $10^{0.89}$ \\
23 & 2.300 & 1.800 &-0.150&-0.150 & 0.5 & 63.700 & $10^{0.95}$ & $10^{0.84}$ \\
24 & 2.400 & 1.900 &-0.200&-0.200 & 0.5 & 65.800 & $10^{0.92}$ & $10^{0.79}$ \\
\hline
\end{tabular}
\end{center}
\label{secondmodel}
\end{table}%

\subsection{$SU(N)$ with $F = 11$ and $A^3$ superpotential}

In this case there are significantly more couplings in the superpotential. The marginal superpotential couplings and vanishing anomalies give us conditions
\beq
2 &=& R_{T_1} + R_{\overline Q_1} + R_{\overline Q_2}  \\ \nonumber
2 &=& R_{T_2} + R_{\overline Q_1} + R_{\overline Q_2} + R_A \\\nonumber
2 &=& R_{F_1} + R_{Q_1} + R_{\overline Q_0} \\\nonumber
2 &=& R_{F_1} + R_{Q_2} + R_{Q_0} \\\nonumber
2 &=& R_{F_2} + R_{Q_1} + R_{\overline Q_0} + R_A \\\nonumber
2 &=& R_{F_2} + R_{Q_2} + R_{Q_0} + R_A \\\nonumber
2 &=& 3 R_A \\\nonumber
0 &=& N + \frac{5}{2} (R_{Q_1} - 1) + \frac{5}{2} (R_{Q_2} - 1) + \frac{5}{2} (R_{\overline Q_1} - 1) + \frac{5}{2} (R_{\overline Q_2} - 1)  \\\nonumber
& &+ \frac{1}{2} (R_{Q_0} - 1) + \half (R_{\overline Q_0} - 1) + N ( R_A - 1) 
\eeq
In the window of interest several mesons go free: $Q_1 Q_2$, $Q_1 \overline Q_1$, $Q_2 \overline Q_2,\overline Q_1 \overline Q_2, \overline Q_1 Q_0, \overline Q_2 \overline Q_0,$ and $Q_0 \overline Q_0$. The $\overline{\mathbf{10}}$ component of $\overline Q_1 \overline Q_2$ and the linear combination $Q_1 \overline Q_0 + Q_2 Q_0$ are set to zero in the chiral ring.  The resulting $R$-charges are shown below:

\begin{table}[h!!]
\caption{R charges for $SU(N)$ theory with $F=11$ flavors and cubic superpotential}
\begin{center}
\begin{tabular}{|c|c|c|c|c|c|c|c|c||c|c|c|c|}
\hline
$N$ & $R_{T_1}$ & $R_{T_2}$ & $R_{F_1}$ & $R_{F_2}$ & $R_{Q_{1,2}}$ & $R_{\overline Q_{1,2}}$ & $R_{Q_0/\bar Q_0}$ & $R_A$ &  $\mathfrak{b}$ & $\frac{\Lambda_L}{ \Lambda_*}$ & $\frac{\Lambda_T}{\Lambda_*}$ & $ \frac{\Lambda_F}{\Lambda_*}$ \\ \hline
11 & 1.333 & 0.833 & 1.333 & 0.667 & 0.333 & 0.333 & 0.333 & 0.667 & 36.665 & $10^{1.65}$ & - & -  \\  
12 & 1.452 & 0.952 & 1.433 & 0.766 & 0.266 & 0.274 & 0.301 & 0.667 & 36.098 & $10^{1.67}$ & $10^{2.14}$ & - \\
13 & 1.578 & 1.078 & 1.529 & 0.862 & 0.202 & 0.211 & 0.269 & 0.667 & 36.674 & $10^{1.64}$ & $10^{1.49}$ & $10^{1.78}$ \\
14 & 1.711 & 1.211 & 1.623 & 0.956 & 0.142 & 0.145 & 0.236 & 0.667 & 37.976 & $10^{1.59}$ & $10^{1.31}$ & $10^{1.20}$ \\
15 & 1.851 & 1.351 & 1.715 & 1.048 & 0.085 & 0.075 & 0.200 & 0.667 & 40.002 & $10^{1.51}$ & $10^{1.15}$ & $10^{0.91}$ \\
16 & 1.995 & 1.495 & 1.806 & 1.140 & 0.032 & 0.002 & 0.162 & 0.667 & 42.758 & $10^{1.41}$ & $10^{1.02}$ & $10^{0.74}$ \\ \hline
\end{tabular}
\end{center}
\label{thirdmodel}
\end{table}

\vspace{1mm}

\subsection{$SU(N)$ with $F=10$, $A^3$ superpotential, and marginal coupling to $\Sigma$}

Very little changes from the simple case of \S A.1 if we add couplings to the adjoint Higgs $\Sigma$ of $SU(5)_{SM}$; only the contribution to $\mathfrak{b}$ is modified. The R-charges for additional coupling to the adjoint Higgs of $SU(5)$ are given below:

\begin{table}[h!!]
\caption{R charges for $SU(N)$ theory with $F=10$ flavors, cubic superpotential, and coupling to $SU(5)$ adjoint}
\begin{center}
\begin{tabular}{|c|c|c|c|c|c|c||c|c|c|}
\hline
$N$ & $R_{T_1}$ & $R_{T_2}$ & $R_{Q_{1,2}}$ & $R_{\overline Q_{1,2}}$ & $R_A$ & $R_{\Sigma}$ &$\mathfrak{b}$ & $\frac{\Lambda_{L}}{ \Lambda_*}$ & $\frac{\Lambda}{ \Lambda_*}$ \\ \hline
11 & 1.448 & 0.781 & 0.257 & 0.276 & 0.667 & 1.467 & 23.884 & $10^{2.52}$ & -  \\  
12 & 1.572 & 0.905 & 0.186 & 0.214 & 0.667 & 1.600 & 24.528 & $10^{2.46}$ & - \\
13 & 1.705 & 1.039 & 0.119 & 0.147 & 0.667 & 1.733 & 25.925 & $10^{2.32}$ & $10^{1.649}$ \\
14 &  1.849 & 1.182 & 0.058 & 0.076 & 0.667 & 1.867 & 28.080 & $10^{2.15}$ & $10^{1.190}$ \\ \hline
\end{tabular}
\end{center}
\label{fourthmodel}
\end{table}%

\subsection{$Sp(2N)$ with $2F=10$ and $A^3$ superpotential}

In this case the constraints from superpotential couplings and anomalies are 
\beq
2 &=& R_{T_1} + 2 R_{\bar Q} \\ \nonumber
2 &=& R_{T_2} + 2 R_{\bar Q} + R_A \\ \nonumber
0 &=& 2 (N+1) + 5 (R_Q -1) + 5(R_{\bar Q} - 1) + 2 (N-1)(R_A - 1) \\ \nonumber
2 &=& 3 R_A
\eeq
Now the operators that go free are the mesons $QQ$ and $Q \overline Q$; the meson $\overline Q \overline Q$ is set entirely to zero in the chiral ring, and there are no baryons. The resulting $R$-charges are given below:

\begin{table}[h!!]
\caption{R charges for $Sp(2N)$ theory with $2F = 10$ flavors and cubic antisymmetric superpotential}
\begin{center}
\begin{tabular}{|c|c|c|c|c|c||c|c|c|}
\hline
$N$ & $R_{T_1}$ & $R_{T_2}$ & $R_{Q}$ & $R_{\overline Q}$ & $R_A$  & $\mathfrak{b}$ & $\frac{\Lambda_{L}}{\Lambda_*}$ & $\frac{\Lambda}{ \Lambda_*}$ \\ \hline
4 & 1.497& 0.830 & 0.149 & 0.251 & 0.667 & 6.06 & $10^{9.94}$ & - \\
5 &  1.786 & 1.119 & 0.026 & 0.107 & 0.667 & 7.16 & $10^{8.41}$ & $10^{1.36}$ \\ \hline
\end{tabular}
\end{center}
\label{sp2n-ten}
\end{table}%

\subsection{$Sp(2N)$ with $2F=10$ and $A^4$ superpotential}

The constraints in this case are a simple generalization of the previous case,
\beq
2 &=& R_{T_1} + 2 R_{\bar Q} \\ \nonumber
2 &=& R_{T_2} + 2 R_{\bar Q} + R_A \\ \nonumber
0 &=& 2 (N+1) + 5 (R_Q -1) + 5(R_{\bar Q} - 1) + 2 (N-1)(R_A - 1) \\ \nonumber
2 &=& 4 R_A
\eeq
The operators that can go free are now $QQ$, $Q \overline Q$, and for sufficiently high $N$ both $QAQ$ and $Q A \overline Q$. The resulting $R$-charges are shown below.

\begin{table}[h!!]
\caption{R charges for $Sp(2N)$ theory with $2F = 10$ flavors and quartic antisymmetric superpotential}
\begin{center}
\begin{tabular}{|c|c|c|c|c|c||c|c|c|}
\hline
$N$ & $R_{T_1}$ & $R_{T_2}$ & $R_{Q}$ & $R_{\overline Q}$ & $R_A$  & $\mathfrak{b}$ & $\frac{\Lambda_{L}}{ \Lambda_*}$ & $\frac{\Lambda}{ \Lambda_*}$ \\ \hline
4 & 1.331 & 0.831 & 0.266 & 0.334 & 0.500 & 8.46 & $10^{7.12}$ & - \\
5 & 1.531 & 1.031 & 0.166 & 0.234 & 0.500 & 9.96 & $10^{6.05}$ & $10^{1.68}$ \\
6 &  1.787 & 1.287 & 0.093 & 0.107 & 0.500 & 12.41 & $10^{4.86}$ & $10^{1.22}$ \\
7 & 2.000 & 1.500 & 0.000 & 0.000 & 0.500 &  13.00 & $10^{4.64}$ & $10^{1.02}$ \\
8 & 2.200 & 1.700 & -0.100 & -0.100 & 0.500 & 14.20 & $10^{4.24}$ & $10^{0.89}$ \\
\hline
\end{tabular}
\end{center}
\label{fifthmodel}
\end{table}%

\subsection{$Sp(2N)$ with $2F = 12$ and $A^3$ superpotential}

The constraints from marginal superpotential terms and anomaly cancellation are
\beq
2 &=& R_{T_1} + 2 R_{\bar Q} \\ \nonumber
2 &=& R_{T_2} + 2 R_{\bar Q} + R_A \\ \nonumber
2 &=& R_{F_1} + R_{Q} + R_{Q_0} \\ \nonumber
2 &=& R_{F_1} + R_{Q} + R_{\bar Q_0} \\ \nonumber
2 &=& R_{F_2} + R_{Q} + R_A + R_{Q_0} \\ \nonumber
2 &=& R_{F_2} + R_{Q} + R_A + R_{\bar Q_0} \\ \nonumber
2 &=& 3 R_A\\ \nonumber
0 &=& 2 (N+1) + 5 (R_Q -1) + 5(R_{\bar Q} - 1) + (R_{Q_0} - 1) + (R_{\bar Q_0} - 1) + 2 (N-1)(R_A - 1) 
\eeq

The gauge-invariant chiral operators  set to zero in the chiral ring are $\overline Q \overline Q, \overline Q A \overline Q,$ and the linear combinations $Q Q_0 + Q \overline Q_0, Q A Q_0 + Q A \overline Q_0$. Of the remaining chiral operators, $QQ, Q \overline Q, \overline Q Q_0, \overline Q \overline Q_0,$ and $Q_0 \overline Q_0$ saturate the unitarity bound and must be accounted for accordingly in the $a$-maximization procedure. The resulting $R$-charges are shown below.

\begin{table}[h!!]
\caption{R charges for $Sp(2N)$ theory with $2F=12$ flavors and cubic superpotential}
\begin{center}
\begin{tabular}{|c|c|c|c|c|c|c|c|c||c|c|c|c|}
\hline
$N$ & $R_{T_1}$ & $R_{T_2}$ & $R_{F_1}$ & $R_{F_2}$ & $R_{Q}$ & $R_{\overline Q}$ & $R_{Q_0/\bar Q_0}$ & $R_A$ &  $\mathfrak{b}$ & $\frac{\Lambda_L}{ \Lambda_*}$ & $\frac{\Lambda_T}{\Lambda_*}$ & $ \frac{\Lambda_F}{\Lambda_*}$ \\ \hline
4 & 1.333 & 0.667 & 1.333 & 0.667 & 0.333 & 0.333 & 0.333 & 0.667 & 9.000 & $10^{6.70}$ & - & -  \\  
5 & 1.549 & 0.882 & 1.523 & 0.856 & 0.195 & 0.225 & 0.282 & 0.667 & 8.483 & $10^{7.10}$ & - & - \\
6 & 1.847 & 1.181 & 1.674 & 1.007 & 0.100 & 0.076 & 0.227 & 0.667 & 9.120 & $10^{6.61}$ & $10^{1.19}$ & $10^{1.02}$ 
 \\ \hline
\end{tabular}
\end{center}
\label{spgenmodel}
\end{table}

%%%%%%%%%%%%%%%%%%%
\vspace{1cm}
\pagebreak

\bibliography{scftflavorbib}
\bibliographystyle{JHEP}

\end{document}